\documentstyle[12pt]{article}
\def\subsubsection{\@startsection{subsubsection}{3}{\z@}{-3.25ex plus
 -1ex minus -.2ex}{1.5ex plus .2ex}{\large\sc}}

\newcommand{\be}{\begin{equation}}
\newcommand{\bel}[1]{\begin{equation}\label{#1}}
\newcommand{\ee}{\end{equation}}
\newcommand{\bea}{\begin{eqnarray}}
\newcommand{\ba}{\begin{array}}
\newcommand{\eea}{\end{eqnarray}}
\newcommand{\ea}{\end{array}}
\newcommand{\noin}{\noindent}
\newcommand{\ra}{\rightarrow}

\newcommand{\hfour}{\hspace*{4mm}}

\newcommand{\vtwo}{\vspace*{2mm}}
\newcommand{\htwo}{\hspace*{2mm}}
\newcommand{\honecm}{\hspace*{1cm}}

\newcommand{\htwocm}{\hspace*{2cm}}

\newcommand{\udl}{\underline}

\begin{document}

%%%%%%%%%%%%%%%%%%%%%%%%%%%%%%%
\begin{titlepage}             %
\thispagestyle{empty}         %
\begin{center}                %
\vspace*{1cm}                 %
{\large
%%%%%%%%%%%%%%%%%%%%%%%%%%%%%%%

{\bf
Diffusion-Annihilation in the Presence of a Driving Field
}

%%%%%%%%%%%%%%%%%%%%%%%%%%%%%%%
}\vspace{3cm}                 %
%                              %
{\large {\sc
%%%%%%%%%%%%%%%%%%%%%%%%%%%%%%%

%\author
Gunter M. Sch\"utz
%}
%%%%%%%%%%%%%%%%%%%%%%%%%%%%%%%
}}\\[8mm]                     %
%%%%%%%%%%%%%%%%%%%%%%%%%%%%%%%
%\address
{
Theoretical Physics, University of Oxford,\\
1 Keble Road, Oxford OX1 3NP, UK
}
%%%%%%%%%%%%%%%%%%%%%%%%%%%%%%%
\vspace{5cm}\\                %
%%%%%%%%%%%%%%%%%%%%%%%%%%%%%%%
%\vspace*{-1cm}
%\abstract\begin{center}\hspace*{2cm}
\begin{minipage}{14cm}{\small\sl\rm
We study the effect of an external driving force on a simple stochastic
reaction-diffusion
system in one dimension. In our model each lattice site may be occupied
by at most one particle. These particles hop with rates $(1\pm\eta)/2$ to the
right and left nearest neighbouring site resp. if this site is vacant and
annihilate with rate 1 if it is occupied. We show that density fluctuations
(i.e. the $m^{th}$ moments $\langle N^m \rangle$ of the density distribution
at time $t$) do not depend on the spatial anisotropy $\eta$ induced by the
driving field, irrespective of the initial condition. Furthermore we show
that if one takes certain translationally invariant averages over initial
states
(e.g. random initial conditions) even local fluctuations do not depend on
$\eta$. In the scaling regime $t \sim L^2$ the effect of the driving can be
completely absorbed in a Galilei transformation.
We compute the probability of finding
a system of $L$ sites in its stationary state at time $t$ if it was fully
occupied
at time $t_0 = 0$.
}
\end{minipage}
\end{center}
%\endabstract
%\maketitle]
%%%%%%%%%%%%%%%%%%%%%%%%%%%
\end{titlepage}           %
\baselineskip 0.3in       %
%%%%%%%%%%%%%%%%%%%%%%%%%%%

\section{Introduction}

Stochastic reaction-diffusion processes in one dimension have received
a considerable amount of attention (for a vrief review see e.g. \cite{1}).
Despite their simplicity they show
a very rich behaviour and some of the results obtained so far are of
experimental relevance
\cite{2}. Another reason for their popularity is the wide range of
applicability: Systems of this kind map to interface dynamics \cite{KS},
to polymers in random media \cite{prm} or, quite close to every day life,
to traffic problems \cite{traffic}. This list is by no means exhaustive,
but provides already as it is ample motivation for the study of such models.
Last, but not least, they map to well-known problems in many-body physics,
particularly to integrable vertex models, and by now a considerable amount
of exact and rigorous results have been obtained using this mapping \cite{rdp}.

Here we consider a model defined on a ring of $L$ sites with periodic boundary
conditions where each lattice site may be occupied by at most one particle.
These particles (denoted $A$) hop with rates $(1\pm\eta)/2$ to the
right or left nearest neighbouring site resp. if this site is vacant (denoted
$\emptyset$) and
annihilate with rate $\lambda$ if it is occupied:
\bea
A \emptyset & \ra & \emptyset A \htwocm (1+\eta)/2 \\
\emptyset A & \ra & A \emptyset \htwocm (1-\eta)/2 \\
A A         & \ra & \emptyset \emptyset \htwocm \lambda
\eea
(In this paper we study only $\lambda=1$.)
This model, which is closely related to zero temperaure Glauber dynamics
\cite{Glauber,robin}, has been studied by a number of authors over the past few
years \cite{robin} - \cite{PBG}. In the limit $\lambda=0$ the model reduces
to the well-known asymmetric exclusion process \cite{ZS} whereas for
$\lambda \ra \infty$, equivalent to the absence of any diffusion, the
model becomes equivalent to random sequential adsorption of dimers
\cite{Evans}.
A discrete time version of the model was studied in \cite{Pr1}.
Physically,
the exclusion principle corresponds to a hard-core on-site repulsion and
the asymmetry in the left and  right hopping rates may be thought of as the
result of a field driving the particles in one preferred direction. The
pair annihilation finally takes into a acount the possibility of an effective
short-range attraction leading to unbreakable, inert pairs of particles.

For $\lambda=0$ the effect of the driving field has been well studied
and it turns out to be rather drastic. As opposed to the driven model without
exclusion where the drift may be absorbed in a simple lattice Galilei
transformation \cite{HS2}, here the asymmetry leads to the formation of
shocks \cite{shock}. Not many exact results are known for the time evolution
of the system, but the mapping to the six-vertex model has been shown
to be useful by using Bethe ansatz \cite{GS,S} and related methods \cite{SS}.
Among other things the dynamical exponent appearing in the dynamical
structure function of the system was found to be
$z=3/2$ \cite{GS} rather than $z=2$ in the undriven exclusion process
or the non-interacting driven system without exclusion.
For non-zero, but small annihilation rates $\lambda$ a recent study based on
scaling arguments, mean-field approaches, random walk considerations and
numerical results has given a very different scenario: in the presence
of pair annihilation the effect of the driving appears to be very small
\cite{PBG}. In addition to that, it is known that for $\lambda=1$ and an
initially full lattice neither the time-dependent density nor the
two-point density correlation function depend
on the driving \cite{robin}. These observations make a more detailed study
of the system desirable.

It is the aim of this paper to provide some exact results concerning the
driving
in the presence of the reaction. In Sec. 2 we define the model in terms
of a master equation written in a quantum Hamiltonian formalism.
We shall focus on the case $\lambda=1$
because for that particular choice the model may be described in terms
of free fermions and treated rigourosly. Exact expressions for various
correlation functions
(in principle, for {\em all} correlation functions) become readily available.
It turns out (Sec. 3) that fluctuations in the total density of particles do
not
depend on the driving field at all. The same is true for some
other average values, to be specified below. Furthermore, if one takes certain
translationally invariant averages over initial states, e.g. random initial
conditions, arbitrary {\em local} density correlations show no dependance on
the asymmetry $\eta$. These results generalize
earlier findings \cite{sh}.
The general solution of the master equation and some further results concerning
the scaling regime are given in Sec. 4.
In Sec. 5 we summarize and discuss
our results.

\section{The master equation in quantum Hamiltonian formulation}
\setcounter{equation}{0}

\subsection{Definitions}

We define the process in terms of a master equation for the probability
$f(\underline{n};t)$ of finding, at time $t$, any configuration $\underline{n}$
of particles in the system of $L$ sites with periodic boundary conditions. Here
$\udl{n} = \{n_1, n_2, \dots , n_L\}$ where $n_i = 0,1$ and
$1 \leq i \leq L$ labels the sites of the lattice.
An alternative possibility is to give the set
$\{ x_1, x_2,
\dots , x_N \}$ of occupied lattice sites. In this notation,
the empty set
represents the empty lattice and $1 \leq N \leq L$ is the total number of
particles in the configuration.
We shall express the time evolution
given by the master equation in terms of a quantum Hamiltonian $H$
\cite{qhf}.
This is  discussed in detail in a number of publications (for consistency
of notation see e.g. \cite{rdp})
and we shall repeat only the essential elements of the mapping.
The advantage of this approach is that there are standard methods of dealing
with the resulting time evolution operator $H$. The applicability of these
techniques, in the case
at hand essentially a Jordan Wigner transformation and the representation
of states in terms of a fermionic Fock space, does not arise naturally
and obviously if the master equation is written down in standard form.

The idea is to represent each of the $2^L$ possible configurations
in $X=\{0,1\}^L$ by a
vector $|\, \udl{n} \,\rangle$ (or $|\, x_1, \dots , x_N \,\rangle$, with
$|\,0\,\rangle\equiv |\;\;\rangle$ being the empty state). The probability
distribution is then mapped to a state vector
\bel{2-1}
| \, f(t)\, \rangle = \sum_{\udl{n} \in X} f(\udl{n};t)
|\, \udl{n} \,\rangle \htwo .
\ee
The vectors $|\, \udl{n} \,\rangle$
together with the
transposed vectors $\langle \, \udl{n} \, |$ form an orthonormal basis
of $(C^2)^{\otimes L}$ and the time evolution is defined
in terms of a linear 'Hamilton' operator $H$ acting on this space of
dimension $2^L$
\bel{2-2}
\frac{\partial}{\partial t} | \, f(t)\, \rangle =
- H | \, f(t)\, \rangle \htwo .
\ee
A state at time $t=t_0 + \tau$ is therefore given in terms of an initial
state at time $t_0$ by
\bel{2-2a}
| \, f(t_0+\tau \, \rangle = \mbox{e}^{-H\tau }
| \, f(t_0 \, \rangle \htwo .
\ee
{}From (\ref{2-1}) and (\ref{2-2}) and using
$f(\udl{n};t) = \langle \, \udl{n} \, | \, f(t) \, \rangle$
the master equation takes the form
\bel{2-3}
\frac{\partial}{\partial t} f(\udl{n};t) = - \langle \, \udl{n} \, |
H | \, f(t) \, \rangle \htwo .
\ee
Note that
\bel{2-4}
\langle \, s \,|\, f(t) \, \rangle = \sum_{\udl{n} \in X} f(\udl{n};t) = 1
\ee
where
\bel{2-5}
\langle \, s \,| = \sum_{\udl{n} \in X} \langle \, \udl{n} \, |
\ee
which expresses conservation of probability and which implies
$\langle \, s \,| H = 0$.

Expectation values $\langle \, Q \, \rangle$ are calculated as matrix elements
of suitably chosen operators $Q$.
A complete set of observables are the occupation numbers $n_k=0,1$.
Defining projection {\em operators} on states with a particle on site $k$ of
the chain as
\bel{2-7}
n_k = \frac{1}{2} \left( 1 - \sigma^z_k \right) =
\left(
\ba{cc}
0 & 0 \vtwo \\
0 & 1
\ea
\right)_k
\ee
one finds that the average density of particles at
site $k$ is given by $\langle \, n_k \, \rangle =
\langle \, s \,| n_k | \, f(t) \, \rangle$.
Correlation functions
$\langle \, n_{k_1} \cdots n_{k_j} \, \rangle$, i.e., the
probabilities of finding particles on the set of sites
$\{k_1, \dots ,k_j\}$,
are computed analogously.

For later convenience we also introduce the operators
$s^{\pm}_k = (\sigma^x_k \pm i \sigma^y_k)/2$.
In our convention
\bel{2-8}
s^-_k =
\left(
\ba{cc}
0 & 0 \vtwo \\
1 & 0
\ea
\right)_k
\ee
creates a particle at site $k$ when acting to the right, while
\bel{2-9}
s^+_k =
\left(
\ba{cc}
0 & 1 \vtwo \\
0 & 0
\ea
\right)_k
\ee
annihilates a particle at site $k$. Note that
\bel{2-10}
\langle \, s \,| s^+_k = \langle \, s \,| n_k
\hfour \mbox{and}  \hfour
\langle \, s \,| s^-_k = \langle \, s \,| (1-n_k ) \htwo .
\ee
Introducing the ladder operators $S^{\pm} = \sum_{k=1}^L s^{\pm}_k$
one may write
\bel{2-10a}
\langle \, s \,| = \langle \, 0 \,| \, \mbox{e}^{S^+} \htwo .
\ee
Using the commutation relations for the Pauli matrices then yields
(\ref{2-10}).

Now we can define the process in terms of the quantum Hamiltonian
\bel{2-11}
H = \sum_{k=1}^L u_k
\ee
with the nearest neighbour reaction matrices
\bea
u_k & = & \frac{1+\eta}{2} \left( n_k (1-n_{k+1}) - s^+_k s^-_{k+1} \right) +
          \frac{1-\eta}{2} \left( (1-n_k) n_{k+1} - s^-_k s^+_{k+1} \right)
          \nonumber \\
    &   & + \lambda \left( n_k n_{k+1} - s^+_k s^+_{k+1} \right) \\
    & \stackrel{\lambda\rightarrow 1}{=} &  \frac{1+\eta}{2}
          (n_k - s^+_k s^-_{k+1}) - \frac{1-\eta}{2}( n_{k+1} - s^-_k
s^+_{k+1})
          \nonumber
\eea
This, together with (\ref{2-3}) defines the process. The
Hamiltonian may be written $H=H_s + \eta H_d$ where the driving part $H_d$
is given by $H_d=1/2\sum_{k=1}^{L} (s^-_k s^+_{k+1} - s^+_k s^-_{k+1})$
and $H_s$ is the Hamiltonian for the system without driving.

\subsection{Fermion representation of $H$}

For $\lambda=1$ the Hamiltonian becomes bilinear in the creation and
annihilation
operators $s^{\pm}_k$. This suggests rewriting $H$ by introducing
fermionic operators through a Jordan-Wigner transformation \cite{JW}. We define
\bea
\label{2-12} Q_k & = & \prod_{i=1}^{k} \sigma^z_i \\
\label{2-13} a^{\dagger}_k & = & s^{-}_k Q_{k-1} \\
\label{2-14} a_k & = & Q_{k-1} s^{+}_k
\eea
satisfying the anticommutation relations $\{a_k,a_l\}=\{a^{\dagger}_k,
a^{\dagger}_l\}=0$ and $\{a^{\dagger}_k,a_l\}=\delta_{k,l}$. Note that
because of the periodic boundary conditions for the Pauli matrices
one has $a^{\dagger}_{L+1} = a^{\dagger}_1 Q_L$ and $a_{L+1} = Q_L a_1$.
$Q_L$ may be written $Q_L=(-1)^{N}$ where $N=\sum n_k$ is the number operator.
Since by the action of $H$ the particle number changes only in units of two,
$Q_L$ commutes with
$H$ and splits it into a sector with an even number of particles ($Q_L=+1$)
and into a sector with an odd number of particles ($Q_L = -1$).
In terms of the fermionic operators one finds
\bea
s^{+}_k s^{-}_{k+1} \; = \; a^{\dagger}_{k+1} a_k &  &
s^{-}_k s^{+}_{k+1} \; = \; a^{\dagger}_k a_{k+1}  \\
s^{+}_k s^{+}_{k+1} \; = \; a_{k+1} a_k &  &
s^{-}_k s^{-}_{k+1} \; = \; a^{\dagger}_k a^{\dagger}_{k+1} \\
n_k & = & a^{\dagger}_k a_k
\eea
and we arrive at the following expressions for $H_s$ and $H_d$:
\bea
\label{2-15}
H_s & = & - \frac{1}{2} \sum_{k=1}^L \left\{ a^{\dagger}_{k+1} a_k +
          a^{\dagger}_k a_{k+1} + 2 a_{k+1} a_k - 2 n_k \right\} \nonumber \\
    &   & - (\lambda-1) \sum_{k=1}^L \left\{a_{k+1} a_k - n_k n_{k+1}\right\}
\\
\label{2-16}
H_d & = & - \frac{1}{2} \sum_{k=1}^L \left\{ a^{\dagger}_{k+1} a_k -
          a^{\dagger}_k a_{k+1} \right\}
\eea
We also note that $H_s$ and $H_d$ commute if $\lambda=1$.
Independently of $\lambda$ and $\eta$ the non-degenerate ground state of $H$,
corresponding to the steady state of the system, is the totally empty state
$| 0 \rangle$ in the sector with an even number of particles and the state
$| 0 \rangle^{odd} = 1/L \sum_{k=1}^{L} | k \rangle$ where one particle may
be found with equal probability $1/L$ anywhere in the lattice in the sector
with an odd number of particles.

Since expectation values $\langle n_k(t) \rangle$
of the stochastic variables $n_k=0,1$ are given
by the matrix elements $\langle s | n_k \exp{(-Ht)} | f \rangle$
of the operator $n_k$ and since also $\langle s | \exp{(Ht)} = \langle s |$,
one may introduce time-dependent operators
\bel{2-17}
{\cal O}(t)= \exp{(Ht)} {\cal O}
\exp{-(Ht)}
\ee
and write $\langle {\cal O}(t) \rangle =
\langle s | {\cal O}(t) | f \rangle$ for an arbitrary operator ${\cal O}$.
According to the definition (\ref{2-17})
one has
\bel{2-18}
\frac{d}{dt} {\cal O} \equiv \dot{{\cal O}} = \left[ H , {\cal O} \right]
\ee
Among the quantities of interest are the density of particles in the system and
fluctuations in this quantity. The $m^{th}$ moment of the particle number
distribution is the expectation value $\langle N^m (t) \rangle$ where
$N$ is the number operator $N=\sum_{k=1}^{L} a^{\dagger}_k a_k$.
Any $\cal{ O}$ may be written as a product of the fermionic annihilation and
and creation operators and it is therefore sufficient to study the time
evolution of these operators.

It is useful to introduce the Fourier transforms
\bea
\label{2-19}
b_p & = & \frac{e^{-i\frac{\pi}{4}}}{\sqrt{L}}
\sum_{k=1}^{L} e^{\frac{2\pi i k p}{L}} a_k \\
\label{2-20}
b^{\dagger}_p & = & \frac{e^{i\frac{\pi}{4}}}{\sqrt{L}}
                    \sum_{k=1}^{L} e^{-\frac{2\pi i k p}{L}} a^{\dagger}_k
\eea
satisfying $\{b_p,b_q\}=\{b^{\dagger}_p,
b^{\dagger}_q\}=0$ and $\{b^{\dagger}_p,b_q\}= \delta_{p,q}$. Inverting
(\ref{2-19}), (\ref{2-20}) yields
\bea
a_k & = & \frac{e^{i\pi/4}}{\sqrt{L}} \sum_p
e^{\frac{-2 \pi i kp}{L}} b_p \\
a^{\dagger}_k & = & \frac{e^{-i\pi/4}}{\sqrt{L}} \sum_p
e^{\frac{2 \pi i kp}{L}} b^{\dagger}_p
\eea
Thus the representation of the  number operator in Fourier space is
\bel{2-21}
N =  \sum_p b^{\dagger}_p b_p
\ee
Here the sum runs over all integers $p=0, \dots, L-1$ in the sector with an
odd number particles and over the half odd integers $p=1/2, 3/2 \dots L-1/2$
in the even sector.

The Hamiltonian for $\lambda=1$ reads
\bea
\label{2-22}
H_s & = &  \sum_p \left\{ \left(1-\cos{\frac{2\pi p}{L}}\right)
           b^{\dagger}_p b_p
           + \sin{\frac{2\pi p}{L}} b_{-p} b_p \right\} \\
\label{2-23}
H_d & = &  -i \sum_p \sin{(\frac{2\pi p}{L})}
           b^{\dagger}_p b_p
\eea
$H_s$ was already obtained in \cite{ADHR}.

\subsection{Initial states in the fermion representation}

Having discussed the representation of operators  in terms of the fermionic
operators we turn to the representation of initial configurations or states.
In what follows we shall use the term '(initial) configuration' for a vector
$|k_1 , \dots , k_N\rangle$ or $|\underline{n}\rangle$, i.e., for a
simultaneous eigenstate of all $n_k$. (Initial) states $|f\rangle=
\sum_{\underline{n}}f(\underline{n})
|\underline{n}\rangle$ are vectors which may be superpositions of such
configurations, and normalized such that $\langle s | f \rangle =1$.
The vector $| 0 \rangle$ is the vacuum state with respect to the annihilation
operators, $a_k | 0 \rangle = 0$. In spin language this is the ferromagnetic
state with all spins up. Acting with creation operators yields
\bel{2-24}
a^{\dagger}_{k_1} \cdots a^{\dagger}_{k_N} | 0 \rangle =
| k_1, \dots , k_N \rangle  \honecm (k_1 < k_2 < \dots < k_N)
\ee

A general translationally invariant $N$-particle state is obtained by acting
with products $b^{\dagger}_{p_1} \cdots b^{\dagger}_{p_N}$ on $|0\rangle$ where
$\sum_i p_i =0$. A special class are those build by polynomials in the bilinear
expressions $B^{\dagger}_p= b^{\dagger}_{-p} b^{\dagger}_p$ where
$p=1/2,3/2,\dots ,(L-1)/2$.
Among these particular translationally invariant states are uncorrelated random
initial conditions with an even number of particles. The uncorrelated
random initial state with density $\rho$ is the product measure
\bel{2-25}
| \rho \rangle = \left( \ba{c} 1-\rho \\
                                \rho \ea \right)^{\otimes L}
               = \sum_{N=0}^{L} \rho^N (1-\rho)^{L-N} | N \rangle
\ee
where $|N\rangle$ is the $N$-particle state where each configuration
appears with equal weight. By projection on the sectors with even and odd
particle numbers we obtain
\bel{2-26}
| \rho \rangle^{even(odd)} = \frac{1 \pm Q_L}{1 \pm (1-2\rho)^L} |\rho \rangle
\ee
where we have used $Q_L$ defined in (\ref{2-12}).

{}From the representation (\ref{2-22}), (\ref{2-23}) of $H$ and using
$(B_p)^2=(B^{\dagger}_p)^2=0$ one finds for the left zero energy eigenvector
$\langle s|$ of $H$
\bea
\langle s | & = & \langle 0 |  \prod_{p} \left(
1+ \cot{\left(\frac{\pi p}{L}\right)} b_p b_{-p} \right) +
\langle 0 | b_0 \prod_{p'} \left(
1+ \cot{\left(\frac{\pi p'}{L}\right)} b_{p'} b_{-p'}
\right) \nonumber \\
\label{2-26a}
 & = & \langle 0 | e^{ \sum_{p} \left(
1+ \cot{\left(\frac{\pi p}{L}\right)} b_p b_{-p} \right)} +
\langle 0 | b_0 e^{ \sum_{p'} \left(
1+ \cot{\left(\frac{\pi p'}{L}\right)} b_{p'} b_{-p'}
\right)} \\
 & \equiv & \langle s |^{even} + \langle s |^{odd} \nonumber
\eea
The product and sum resp. over $p,p'$ run over $p=1/2, 3/2, \dots ,(L-1)/2$
(even sector), and over $p'=1, 2, \dots ,L/2-1$ (odd sector).
This in turn implies
\bel{2-27}
\frac{1}{N!} \left( \sum_p \cot{\left(\frac{\pi p}{L}\right)}
B^{\dagger}_p \right)^N =
\sum_{1 \leq k_1 < \dots < k_{2N} \leq L} a^{\dagger}_{k_1} \cdots
a^{\dagger}_{k_{2N}}
\ee
Hence
\bea
\label{2-28}
| 2N \rangle & = & \frac{1}{N!} \left( \sum_p \cot{\left(\frac{\pi
p}{L}\right)}
B^{\dagger}_p \right)^N | 0 \rangle  \\
\label{2-29}
| \rho \rangle^{even} & = & \frac{2}{1+ (1-2\rho)^L}
\prod_p \left( (1-\rho)^2 + \rho^2 \cot{\left(\frac{\pi p}{L}\right)}
B^{\dagger}_p \right) | 0 \rangle
\eea
The completely full lattice is simply given by
\bel{2-30}
|L\rangle = \prod_p  B^{\dagger}_p  | 0 \rangle \htwo .
\ee

In what follows we shall focus on the even sector. For a study of correlation
functions (Sec. 4) it is useful to note
\bel{2-31}
\langle s |^{even} \left(b^{\dagger}_p + \cot{\left(\frac{\pi p}{L}\right)}
b_{-p} \right) = 0
\ee
which may be verified using the momentum space representation (\ref{2-26a})
of $\langle s |^{even}$.

\section{Dynamics of the system on a translationally invariant subspace}

In the last section we have introduced the subspace ${\cal V}$ generated by the
operators $B^{\dagger}_p$.
Together with $B_p=b_{p}b_{-p}$ they satisfy the
algebra of the Pauli matrices, with $[B_p,B^{\dagger}_p]=
b_{-p}b^{\dagger}_{-p} -b^{\dagger}_p b_p  \equiv 2C_p$ playing
the role of the $\sigma^z_p$ matrix. $I_p \equiv b^{\dagger}_p b_p -
b_{-p}b^{\dagger}_{-p}$ commutes with all $B^{\dagger}_p,
B_p,C_p$ and acts as unit operator on this subspace. It satisfies
$I_p X_p = X_p I_p = X_p$ for $X_p=B_p, B^{\dagger}_p, C_p$. These relation
are easy to verify by using the anticommutation relations for $b^{\dagger}_p$
and $b_p$.

This subspace is of interest for three reasons. Firstly, $H$ can be written
in terms of these operators:
\bel{3-1}
H = \sum_p \left(1-\cos{\left(\frac{2\pi}{L}\right)}\right) N_p - 2
\sin{\left(\frac{2\pi}{L}\right)} B_p - i \eta
\sin{\left(\frac{2\pi}{L}\right)}
\left(I_p -1 \right)
\ee
Where we have introduced the number operator $N_p = 1- 2C_p$.
Therefore ${\cal V}$ is an invariant subspace of $H$.
Secondly, a physically important class of initial conditions, namely
random initial conditions, including the steady state and the fully
occupied lattice, are in this subspace. Finally, some physically important
expectation values are given by operators constructed from
$B^{\dagger}_p$ and $B_p$. In particular, using (\ref{2-21}) one gets
\bel{3-2}
N = \sum_p N_p
\ee

Without further calculation we can now
state the following results:\\

\noin {\em 1) The subspace ${\cal V}$ of dimension $2^{L/2}$ generated by
$B^{\dagger}_p$ acting on the vacuum state $|0\rangle$
is an invariant subspace of $H$. On this subspace $H_d=0$, i.e., the
driving has no effect on any correlation function if the system is
at time $t=0$ in an initial state which is contained in ${\cal V}$
(e.g. random initial conditions).}\\

That this correct can be seen by  observing that $I_p$ is the
unit operator on this subspace which gives $H_d=0$. As a result,
the state at time $t$ does not depend on $\eta$ which in turns implies
that no correlation in that state can depend on the driving.\\

\noin {\em 2) The time evolution of operators ${\cal O}$ build from
operators $B^{\dagger}_p$ and $B_p$ (e.g. the density operator)
does not depend on the driving, irrespective of the initial condition.}\\

This is again obvious since $H_d$ commutes with any such operator.
Applying this result to powers $N^m$ of $N$ we find\\

\noin{\em 3) The probability $P(N;t)$ of finding precisely $N$ particles at
time
$t$ in the system does not, for any initial condition, depend on the
driving.}\\

The same applies then of course also for the moments
$\langle N^m(t)\rangle $ of this distribution. For $m=1$ this was shown in
\cite{robin}.
An interesting special case is $P_f(0;t)=\langle 0| f(t) \rangle$ giving the
probability that the system has reached the steady state at time $t$ from an
initial state $|f\rangle$.
To obtain this quantity we note that $H$ (\ref{3-1}) restricted to the
subspace ${\cal V}$ is a sum of $2\times 2$ matrices $h_p$
\be
H = 2 \sum_p \left( \ba{cc} 0 & -\sin{2\pi p/L} \vtwo \\
                   0 & 1- \cos{2\pi p/L} \ea \right)_p
\ee
Computing $\exp{-Ht}=\prod_p \exp{-h_pt}$ and taking the product of the
matrix elements $\langle 0 | B^{\dagger}_p(t) | 0 \rangle$ then gives
for random initial conditions (\ref{2-29}) with an even number of particles
the following exact expression:
\bel{3-3}
P_{\rho}(0;t) = \frac{2}{1+ (1-2\rho)^L}
\prod_p \left[ (1-\rho)^2 + \rho^2 \cot{\left(\frac{\pi p}{L}\right)}
\left(1-e^{-2t(1-\cos{(2\pi p/L))}}\right) \right]
\ee
For an initially full lattice this simplifies to
\bel{3-3a}
P_{1}(0;t) = \prod_p \left(1-e^{-2t(1-\cos{(2\pi p/L))}}\right)
\stackrel{L \rightarrow \infty}{\sim} e^{-L/\sqrt{4\pi t}} \htwo .
\ee

\section{Full solution of the master equation}

The master equation is solved if one has found explicit expressions
for $a^{\dagger}_k(t)$ and $a_k(t)$ or their Fourier transforms. The
knowledge of these quantities allows then the explicit calculation of
any correlation function for any initial condition, including  multi-time
correlators giving
arbitrary conditional probabilities. We will compute $a^{\dagger}_k(t)$ and
$a_k(t)$ and study the effect of the driving on these quantities for late
times. Applications to specific correlation functions of interest will
be given elsewhere. For definiteness we study only the sector with an
even number of particles and we also assume, as in the preceding section,
$L$ even.

It is interesting to study first the differential equation satisfied
by the local density $\langle n_k(t) \rangle$. Differentiating with respect to
time one finds
\bea
\frac{d}{dt} \langle n_k(t) \rangle & = & \frac{1}{2} \left(
\langle n_{k+1}(t) \rangle + \langle n_{k-1}(t) \rangle -2\langle n_k(t)
\rangle
\right) - \frac{\eta}{2} \left( \langle n_{k+1}(t) \rangle -
\langle n_{k-1}(t) \rangle \right) \nonumber \\
 & & - (\lambda + \eta)\langle n_{k-1}(t) n_k(t) \rangle
     - (\lambda - \eta)\langle n_k(t) n_{k+1}(t) \rangle
\eea
In the linear terms one recognizes a lattice Laplacian and lattice derivative
respectively. For $\lambda=0$ the nonlinear term is also a lattice derivative
and the equation is a discrete form of Burgers equation \cite{Burgers}
describing the evolution of shocks for $\eta \neq 0$.
Any $\lambda > 0$ will result in a strong dampening of the amplitude
and the question arises to which extent the nonlinear effects associated
with the driving continue to
play a role. For $\lambda=1$ this question was partially answered in the
last section, the result being, somewhat surprisingly, that, for the
initial conditions considered, there are now such effects at all. Here
we study the form of local correlations for arbitrary initial conditions.

The simple form of $H$ in Fourier space suggests studying $b^{\dagger}_p(t)$
and $b_p(t)$ rather than $a^{\dagger}_k(t)$ and $a_k(t)$.
Applying (\ref{2-18}) gives a set of two coupled ordinary
differential equations\footnote{
Because of the boundary conditions
these equations describe the time evolution of the creation and annihilation
operators only when applied to products with an even number of operators.
This is not a restriction as all expectation values $\langle n_{k_1} \dots
n_{k_N} \rangle$ are of this form.}
\bea
\frac{d}{dt} b^{\dagger}_p(t) & = & \epsilon_p b^{\dagger}_p(t) +
2 \sin{\left(\frac{2 \pi p}{L}\right)} b_{-p}(t) \\
\frac{d}{dt} b_p(t) & = & - \epsilon_p b_p(t)
\eea
solved by
\bea
\label{4-1}
b^{\dagger}_p(t) & = & e^{\epsilon_p t} \left( b^{\dagger}_p +
\cot{\frac{\pi p}{L}}\left( 1 - e^{-(\epsilon_p + \epsilon_{-p})t} \right)
b_{-p} \right) \\
\label{4-2}
b_p(t) & = & e^{-\epsilon_p t} b_p
\eea
with $b^{\dagger}_p(0)=b^{\dagger}_p$, $b_p(0)=b_p$ and
\bel{4-3}
\epsilon_p = 1 - \cos{\frac{2\pi p}{L}} - i \eta \sin{\frac{2\pi p}{L}}
\ee

{}From (\ref{4-1}) and (\ref{4-2}) one obtains
\bel{4-3a}
n_k(t) = \frac{1}{L} \sum_{p,p'} e^{2\pi i k(p-p')/L} b^{\dagger}_p(t)
b_{p'}(t)
\ee
and therefore an explicit expression of any correlation function at time $t$
in terms of correlators in the initial state. In particular,
one may use (\ref{2-31}) to obtain
\bel{4-3b}
\langle n_k(t) \rangle = \frac{1}{L} \sum_{p,p'} e^{2\pi i k(p-p')/L -
(\epsilon_{-p} + \epsilon_{p'})t} \cot{\frac{\pi p}{L}}
\langle b_{p'}b_p \rangle
\ee
This solves the initial value problem for the average density.

Returning to arbitrary time-dependent correlators we note that (\ref{4-1}) -
(\ref{4-3}) demonstrate the impact of the driving on the system in the
scaling regime.
For times $t\gg L^2$ all correlations decay exponentially with correlation
time $\tau = 2L^2/\pi^2$ resulting from slowest mode $p=1/2$. For times
in the scaling regime $t \sim L^2$ and $L$ large, one may approximate
$\epsilon_p$ by
\bel{4-4}
\epsilon_p \approx - \eta \frac{2 \pi i}{L} p + \frac{2 \pi^2}{L^2} p^2
\ee
In other words, the effect of the driving may be absorbed in a Galilei
transformation $r_i \ra r_i +\eta t$ where $r_i = k_i/L$ are the scaled
space coordinates appearing in the correlation function. Apart from
that there are no other effects, unlike in the absence of annihilation where
the driving induces the evolution of shocks. For {\em arbitrary}
translationally
invariant initial conditions the dependence of correlation functions on
$\eta$ vanishes completely in the scaling limit.

\section{Conclusions}

We have studied the effect of driving in a simple reaction-diffusion system
where exclusion particles hop with rates $(1\pm\eta)/2$ to the right and left
respectively if the nearest neighbour sites are empty and which are annihilated
in pairs with rate 1 if the nearest neighbour site is occupied. We obtained the
following results:\\
(1) Certain translationally invariant time-dependent expectation values
including the $m^{th}$ moments $\langle N^m \rangle$ of the particle number
distribution do not depend on the driving parameter $\eta$, regardless of the
initial condition.\\
(2) Arbitrary time-dependent expectation values of occupation numbers
$\langle n_{k_1} \dots n_{k_N} \rangle$ do not depend on $\eta$ if at time
$t=0$ one takes certain translationally invariant averages over initial states,
e.g. uncorrelated random initial conditions.\\
(3) In the scaling regime $t \approx L^2$ the effect of the driving can be
completely absorbed in a Galilei transformation.\\
(4) The probability that the system has reached its steady state (the empty
lattice) at time $t$ from an uncorrelated random initial condition with density
$\rho$ and an even number of particles is given by (\ref{3-3}).\\

These results have been obtained for a annihilation rate $\lambda=1$.
It would be very interesting to study the system for other values of
$\lambda$ in order to understand the transition to the limiting cases
$\lambda=0$ and $\lambda=\infty$. For $\lambda=0$ (the asymmetric exclusion
process) the introduction of driving has a very strong effect, it causes
the evolution of shocks from local inhomogeneities and it is not clear to
which extent these effects survive in the presence of annihilation. Our
calculation shows that for $\lambda=1$ they are completely absent
in the scaling regime. It is interesting to note that
it is not the exclusion principle as such which is responsible for the
non-linear behaviour of the asymmetric exclusion process, but the strength of
the pair interaction between neighbouring particles. In stochastic language
this is the rate of change of a pair of neighbouring particles $\lambda$
compared to the diffusion rates $(1\pm \eta)/2$.
A satisfactory understanding of
this interplay remains an open problem. Intuitively one would expect
the nonlinear behaviour to vanish for late times for any $\lambda \neq 0$
as the system will then be almost empty and therefore effectively
non-interacting. This is supported by various arguments
put forward in Ref. \cite{PBG} and by the known dynamical exponent $z=2$ of the
asymmetric exclusion process in the regime of low (infinitesimal) densities
\cite{HS} (as opposed to $z=3/2$ for finite densities). But a
qualification of the expression 'late times', and, more importantly,
what happens before such late times remains open to debate.

Another interesting open problem is the approach to local equilibrium,
i.e. to a local extended region of empty sites. The first question to
be asked in this context is the probability of reaching a state with, say,
$M$ empty neighbouring sites. The next question then is how this state
further evolves in time. It is of course not stationary as particles
are injected and absorbed by diffusion at the boundaries of this region, but
with time-dependent (vanishing) rates. For this problem the driving will
make a difference as the injection of particles at the left boundary of
this region will be stronger than the loss of particles (if particles move
preferredly to the left). At the right boundary of this region the situation
will be reversed, the loss will exceed the gain. The result of Sec. 4 show
that the problem with a drift may be obtained from the symmetric problem
by a Galilei transformation and therefore gives a partial answer to this
question. Eq. (\ref{3-3}) gives
the probability of finding the system in global equilibrium which does
not depend on the driving.

\section{Acknowledgments}

This work was supported by a grant of the European Community under
the Human Capital and Mobility program.

\bibliographystyle{unsrt}

\begin{thebibliography}{99}

\bibitem{1}
V. Privman, {\em Dynamics of Nonequilibrium Processes: Surface Adsorption,
Reaction-Diffusion Kinetics, Ordering and Phase Separation}
preprint cond-mat 9312079

\bibitem{2}
Mac Donald, Gibbs, Pipler, Biopolymers \underline{6}, 1 (1968);
R. Kopelman, C.S. Li and Z.-Y. Shi, J. Luminescence \underline{45}, 40 (1990);
R. Kroon, H. Fleurent and R. Sprik, Phys. Rev. E \underline{47}, 2462 (1993).

\bibitem{KS}
J. Krug and H. Spohn, in: {\em  Solids far from Equilibrium}, ed. C. Godreche,
Cambridge, Cambridge University Press (1991) and references therein.

\bibitem{prm}
M. Kardar and Y. C. Zhang, Phys. Rev. Lett. \underline{58}, 2087 (1987).

\bibitem{traffic}
K. Nagel and M. Schreckenberg, J. Phys. I, \underline{2}, 2221 (1992);
A. Schadschneider and M. Schreckenberg, J. Phys. A \underline{26}, L679 (1993).

\bibitem{rdp}
I. Peschel. V. Rittenberg and U. Schultze,
Nucl. Phys. B \underline{340}, 655 (1994);
G. Sch\"utz, J. Stat. Phys. (in press);
and references therein.

\bibitem{Glauber}
R. J. Glauber, J. Math. Phys. \underline{4}, 294 (1963).

\bibitem{Lu}
A. A. Lushnikov, Phys. Lett. A {\bf 120}, 135 (1987).

\bibitem{Sp}
J. L. Spouge, Phys. Rev. Lett. {\bf 60}, 871 (1988).

\bibitem{robin}
M. D. Grynberg, T. J. Newman and R. B. Stinchcombe, Phys. Rev. E
\underline{50},
957 (1994); M. D. Grynberg and R.B. Stinchcombe,
Phys. Rev. Lett. \underline{74}, 1242 (1995).

\bibitem{FA}
J. G. Amar and F. Family, Phys. Rev.A {\bf 41}, 3258 (1990);
F. Family and J. G. Amar, J. Stat. Phys. \underline{65}, 1235 (1991).

\bibitem{HS}
M. Henkel and G. M. Sch\"utz,
Physica A \underline{206}, 187 (1994).

\bibitem{ADHR}
F. C. Alcaraz, M. Droz, M. Henkel and V. Rittenberg,
Ann. Phys. (New York) \underline{230}, 250 (1994).

\bibitem{PBG}
V. Privman, E. Burgos and M. D. Grynberg,
Multiparticle Reactions with Spatial Anisotropy, cond-mat 9410103.

\bibitem{ZS}
Zia and B. Schmittmann,
{\em Statistical mechanics of driven diffusive system}, to
appear in {\em Phase Transitions and Critical Phenomena}, eds. C Domb
and J. Lebowitz (Academic, London).

\bibitem{Evans}
J. W. Evans, Rev. Mod. Phys. \underline{65}, 1281 (1993)

\bibitem{Pr1}
V. Privman, J. Stat. Phys. \underline{72}, 845 (1993).

\bibitem{HS2}
M. Henkel and G. M. Sch\"utz,
Int. J. Mod. Phys. B \underline{8}, 3487 (1994).

\bibitem{shock}
B. Derrida, S. A. Janowsky, J. L. Lebowitz and E. R. Speer,
Europhys. Lett. \underline{22}, 651 (1993).

\bibitem{GS}
L.-H. Gwa and H. Spohn,
Phys. Rev. Lett., \underline{68}, 725 (1992),
Phys. Rev. A, \underline{46}, 844 (1992).

\bibitem{S} G. M. Sch\"utz, J. Stat. Phys. \underline{71}, 471 (1993).

\bibitem{SS}
G. M. Sch\"utz, Phys. Rev. E \underline{47}, 4265 (1993);
S. Sandow and G. M. Sch\"utz,
Europhys. Lett. \underline{26}, 7 (1994).

\bibitem{sh}
G. M. Sch\"utz, Oxford preprint.

\bibitem{qhf}
L. P. Kadanoff and  J. Swift, Phys. Rev \underline{165}, 310 (1968);
M. Doi, J. Phys. A \underline{9}, 1465, 1479 (1976);
P. Grassberger and M. Scheunert, Fortschr. Phys. \underline{28}, 547 (1980).

\bibitem{JW}
P. Jordan and E. Wigner, Z. Phys. \underline{47}, 631 (1928).

\bibitem{Burgers}
J. M. Burgers, {\em The non-linear diffusion equation}, Riedel, Boston (1974).

\end{thebibliography}

\end{document}